# Single-layer graphdiyne on Pt(111): Improved catalysis confined under two-dimensional overlayer

Zheng-Zhe Lin* and Xi Chen

*School of Physics and Optoelectronic Engineering, Xidian University, Xi'an 710071, China*

*Corresponding Author. E-mail address:   zzlin@xidian.edu.cn

**Abstract** – In recent years, two-dimensional confined catalysis, i.e. the enhanced catalytic reactions in confined spaces between metal surface and two-dimensional overlayer, makes a hit and opens up a new way to enhance the performance of catalysts. In this work, graphdiyne overlayer was proposed as a more excellent material than graphene or hexagonal boron nitride for two-dimensional confined catalysis. Density functional theory calculations revealed the superiority of graphdiyne overlayer originated from the steric hindrance effect which increases the catalytic ability and lowers the reaction barriers. Moreover, with the big triangle holes as natural gas tunnels, graphdiyne possesses higher efficiency for the transit of gaseous reactants and products than graphene or hexagonal boron nitride. The results in this work would benefit future development two-dimensional confined catalysis.

## 1. Introduction

Since the birth of graphene [1-3], two-dimensional atomic crystals are of considerable interest because of their unique structural and electronic properties. Possible applications of two-dimensional atomic crystals in catalysis are also paid close attention [4]. People have tried to exploit the specialty of graphene for heterogeneous catalysis [5], electrocatalysis [6, 7] and photocatalysis [8]. However, graphene is conventionally considered as chemically inert due to its saturated C-C bonds. So, vacancies, impurities and chemical modifications are widely considered for enhancing the catalytic ability of graphene [9-16] and other two-dimensional materials [17-25]. In chemically modified systems, low-coordinated transition metal atoms are more active than saturated two-dimensional material surfaces to act as catalytic sites. In recent years, scientists keep pursuing the ultimate activity of transition metal catalysts, i.e. the single-atom catalysts anchored to special substrates, which exhibit more superior catalytic ability than conventional metal nanoparticles [26-29]. For example, single Pt or Ir atom embedded in $FeO_x$ surface has shown their ability to transform CO into other molecules [26, 28]. Single Pd atom embedded in graphene has shown remarkable performance in selective hydrogenation of 1, 3-butadiene [15]. Single Fe atom embedded in graphene has been demonstrated as highly efficient catalyst for benzene oxidation [30].

To achieve the synthesis of single-atom catalysts, state-of-the-art techniques have to be employed, which limits the widespread application of single-atom catalysts. People have also explored the opposite side of ultimate catalyst size, i.e. combining two-dimensional materials with "large" catalysts. It is well established that catalysts in confined spaces can have enhanced activity [31, 32]. In recent years, two-dimensional confined catalysis [33-36], which utilizes the confinement of two-dimensional overlayer to enhance the catalytic ability of metal surface, has made a hit. Much effort was devoted to study two-dimensional graphene or hexagonal boron nitride cover on Pt(111) surface [33, 34, 36]. It has been found that gaseous molecules can readily intercalated under the two-dimensional overlayers and the confined space between the ovelayers and the underlying metal acts as nanoreactors. Such novel idea will open a new era of catalysis.

With the rapid development of two-dimensional materials, more and more candidates for two-dimensional confined catalysis have been born. Several decades ago, graphyne and its family (graphdiyne, graphyne-3 *etc*.) were predicted [37-39] to be two-dimensional layered C allotropes with acetylenic C≡C bonds. In 2010, graphdiyne has been successfully synthesized on the surface of copper via a cross-coupling reaction using hexaethynylbenzene [40]. Recently, large segments of graphyne and graphdiyne films have been successfully synthesized [40-45]. From the structure, the big triangle holes in graphdiyne could accommodate small molecules and act as gas tunnels. It can be inferred that with the natural gas tunnels, graphdiyne overlayer may be more suitable than graphene or hexagonal boron nitride for two-dimensional confined catalysis. Theoretical research has confirmed the possibility of small molecules passing through the acetylenic triangle hole of graphdiyne [46, 47]. In addition, a recent report [48] has suggested graphdiyne as catalyst for CO oxidation. Comprehensively considering above situation, further research on graphdiyne-covered systems could benefit future development two-dimensional confined catalysis.

In this work, density functional theory (DFT) calculations were carried out to investigate two-dimensional confined catalysis on graphdiyne-covered Pt(111) surface. Catalytic CO oxidation on graphdiyne-covered Pt(111) surface was systematically studied and compared with pristine Pt(111) surface. The reaction mechanism of CO oxidation was analyzed to reveal the two-dimensional confined catalytic effect of graphdiyne and the advantage of graphdiyne cover to graphene. Due to the steric hindrance of graphdiyne, the barriers and free energy changes of CO oxidation substeps on graphdiyne-covered Pt(111) surface are correspondingly lower than on pristine Pt(111). The catalytic reaction on graphdiyne-covered Pt(111) surface is more thermodynamically favorable and could be easily proceeded at room temperature. This work primarily explores the advantage of porous two-dimensional material to two-dimensional confined catalysis and provides beneficial information for further development of two-dimensional confined catalysis.

## 2. Computational details

DFT calculations were performed using the Vienna *ab initio* simulation package [49-52]. The projector-augmented wave method [53, 54] was used with a kinetic energy cutoff of 400 eV. The generalized gradient approximation of Perdew-Burke-Ernzerhof [55] was employed as the exchange-correlation functional. Grimme's DFT-D2 correction [56] was employed to account for van der Waals interactions (with $C^6$ = 24.67 J nm$^6$ mol$^{-1}$ and $R^{vdW}$ = 1.75 Å chosen for Pt [33]). The Brillouin-zone integration was performed with 2×2×1 Monkhorst-Pack grid [57] and a Gaussian smearing of $\sigma$ = 0.05 eV. The convergence of total energy was considered to be achieved until the energy difference of two iterated steps was less than 10$^{-6}$ eV. Geometries were fully relaxed without any symmetric constrains until the Hellmann-Feynman forces were below 0.001 eV/Å.

Graphdiyne-covered Pt(111) surface was simulated by a repeated slab model in which (1×1) graphdiyne layer was placed on top of four-layered ($2\sqrt{3} \times 2\sqrt{3}$) Pt slab with the bottom two layers fixed. The replicas of simulation system were separated by a vacuum layer of at least 12 Å in the direction perpendicular to the Pt(111) surface. The size of two fixed bottom layers was set according to the optimized lattice constant $a$=3.97 Å of Pt bulk, which is about 1% larger than the experimental value $a$=3.92 Å. The graphdiyne lattice constant was adapted accordingly, resulting in a strain of 3%.

The search of reaction paths and transition states was performed using the climbing image nudged elastic band (CINEB) method [58-60], with linear interpolation between the coordinates of reactant and product as initial guess of reaction paths. Seven images were inserted between two stable states. The reaction paths were relaxed by minimizing the residual forces with quasi-Newton algorithm. The geometries of reactants, products and transition states were verified by means of frequency calculations. In free energies calculations, the zero-point energy (ZPE) and entropic corrections have been included. At 0 K, the free energy of a species is calculated according to

$$G_0 = E_{DFT} + E_{ZPE}, \tag{1}$$

where $E_{DFT}$ is the relaxed DFT total energy and $E_{ZPE}$ is the ZPE. At $T$ = 300 K and pressure $p$ = 1 atm, the free energy of a species is calculated according to

$$G = E_{DFT} + E_{ZPE} + \int_0^T C_p dT - TS, \tag{2}$$

where $\int_0^T C_p dT$ is the integrated heat capacity, $T$ is the temperature, and $S$ is the entropy. ZPE is calculated with the vibrational frequencies as calculated within DFT.

For gas phase species (CO, $O_2$ and $CO_2$), the integrated heat capacity $\int_0^T C_p dT$ and entropy $S$ are obtained from standard tables of thermodynamic data [61, 62].

For the adsorption of species A on another species B, the binding energy is defined as

$$E_b = ( G_0(A) + G_0(B) - G_0(A\text{-}B) ), \quad (3)$$

where $G_0(A)$, $G_0(B)$ and $G_0(A\text{-}B)$ are the free energy of A, B and A-B complex at 0 K, respectively, including ZPE corrections. For a reaction, the potential barrier is calculated according to

$$E_a = G_0^{\neq} - G_{0\ \text{reactant}}, \quad (4)$$

where $G_{0\ \text{reactant}}$ is the sum of free energies of the reactants at 0 K, and $G_0^{\neq}$ is the free energy of transition state at 0 K. The standard free energy of activation reads

$$\Delta G^{\neq} = G^{\neq} - G_{\text{reactant}}, \quad (5)$$

where $G_{\text{reactant}}$ is the sum of free energies of the reactants and $G^{\neq}$ is the free energy of transition state at $T = 300$ K and pressure $p = 1$ atm. The standard free energy change of a reaction reads

$$\Delta G = G_{\text{product}} - G_{\text{reactant}}, \quad (6)$$

where $G_{\text{product}}$ is the sum of free energies of the products at $T = 300$ K and pressure $p = 1$ atm.

## 3. Results and discussion

### 3.1 *Graphdiyne-covered* Pt(111) *surface*

To obtain the most stable structure of graphdiyne-covered Pt(111) surface, we considered high-symmetry **T**, **B**, **F** and **H** configurations, in which the center of graphdiyne hexagonal is positioned above the surface Pt atom, the bridge of the surface Pt-Pt bond, the FCC hollow site and the HCP hollow site, respectively. The binding energy per C atom of graphdiyne on Pt(111) surface is defined as $E_b/N = ( G_0(\text{Pt}) + G_0(\text{G}) - G_0(\text{Pt-G}) )/N$, where $G_0(\text{Pt})$, $G_0(\text{G})$ and $G_0(\text{Pt-G})$ are the free energy of clean Pt slab, pristine graphdiyne sheet and graphdiyne-covered Pt slab at 0 K, respectively, and $N = 18$ is the number of C atoms in the graphdiyne sheet. The **T** configuration (Fig. 1) was found to be the most stable, with a binding energy $E_b/N$ = 0.091 eV/atom and a distance $d$ = 3.1 Å from the Pt(111) surface to the graphdiyne sheet. The calculated values are in agreement with Ref. [63] ($E_b/N$ = 0.11 eV/atom and $d$ = 2.88 Å). Compared with the binding energy of graphene on Pt (0.042~0.084 eV/atom) [64], graphdiyne is bound more strongly on Pt. It is worth noting that the graphdiyne-Pt distance is close to graphene-Pt and hexagonal-boron-nitride-Pt distance (>3 Å) [33, 34, 64], providing appropriate environment for two-dimensional confined catalysis underneath graphdiyne.

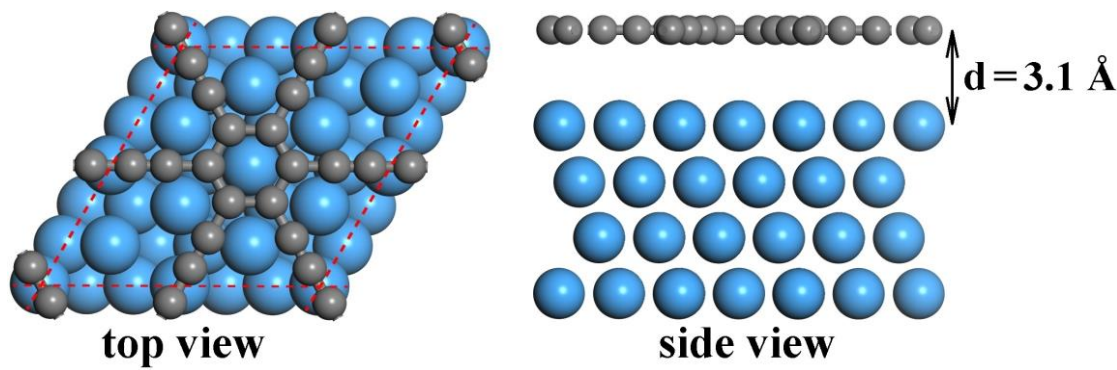

Fig. 1 The top and side view of the **T** configuration of graphdiyne-covered Pt(111) slab. The simulation cell is enclosed by dashed lines. Pt and C atoms are represented in blue and gray, respectively.

3.2 $O_2$ *adsorption and dissociation on graphdiyne-covered* Pt(111) *surface*

Before investigating $O_2$ adsorption and dissociation on graphdiyne-covered Pt(111) surface, we first considered $O_2$ adsorption and dissociation on pristine Pt(111) surface for comparison. Fig. 2(a) exhibits the high-symmetry FCC hollow site, the HCP hollow site, and the BRIDGE site for $O_2$ on Pt(111). We also considered the ATOP site where $O_2$ is positioned on top of surface Pt atom, but the $O_2$ molecule gradually moves to nearby locations during geometry relaxation. The binding energies of $O_2$ on FCC/HCP/BRIDGE sites were found to be $E_b$ = 0.68/0.45/0.69 eV. For the most stable binding on the BRIDGE site, the calculated binding energy is close to reported values in Ref. [65], [66] and [67] (0.69, 0.62 and 0.81 eV, respectively). In the dissociation process of $O_2$ on Pt(111) (i.e. the O-O bond breakage), the O-O bond gradually rotates to the x direction, reaching the transition state with $E_a$ = 0.38 eV and $\Delta G^{\neq}$ = 0.33 eV via a configuration near the FCC site. This calculated reaction path is in agreement with reported in Ref. [67]. The standard free energy change of the dissociation reaction $O_2 \rightarrow 2O$ is $\Delta G$ = -1.62 eV. The binding energies of single dissociated O atom on the FCC/HCP hollow sites are $E_b$ = 4.50/3.96 eV, respectively. Note that on the most stable FCC site, the binding energy of O atom is in agreement with Ref. [65-68] (4.0~4.4 eV). The above calculations reproduced $O_2$ binding energy and dissociation barrier those are similar to previous reports, indicating the reliability of our calculation method.

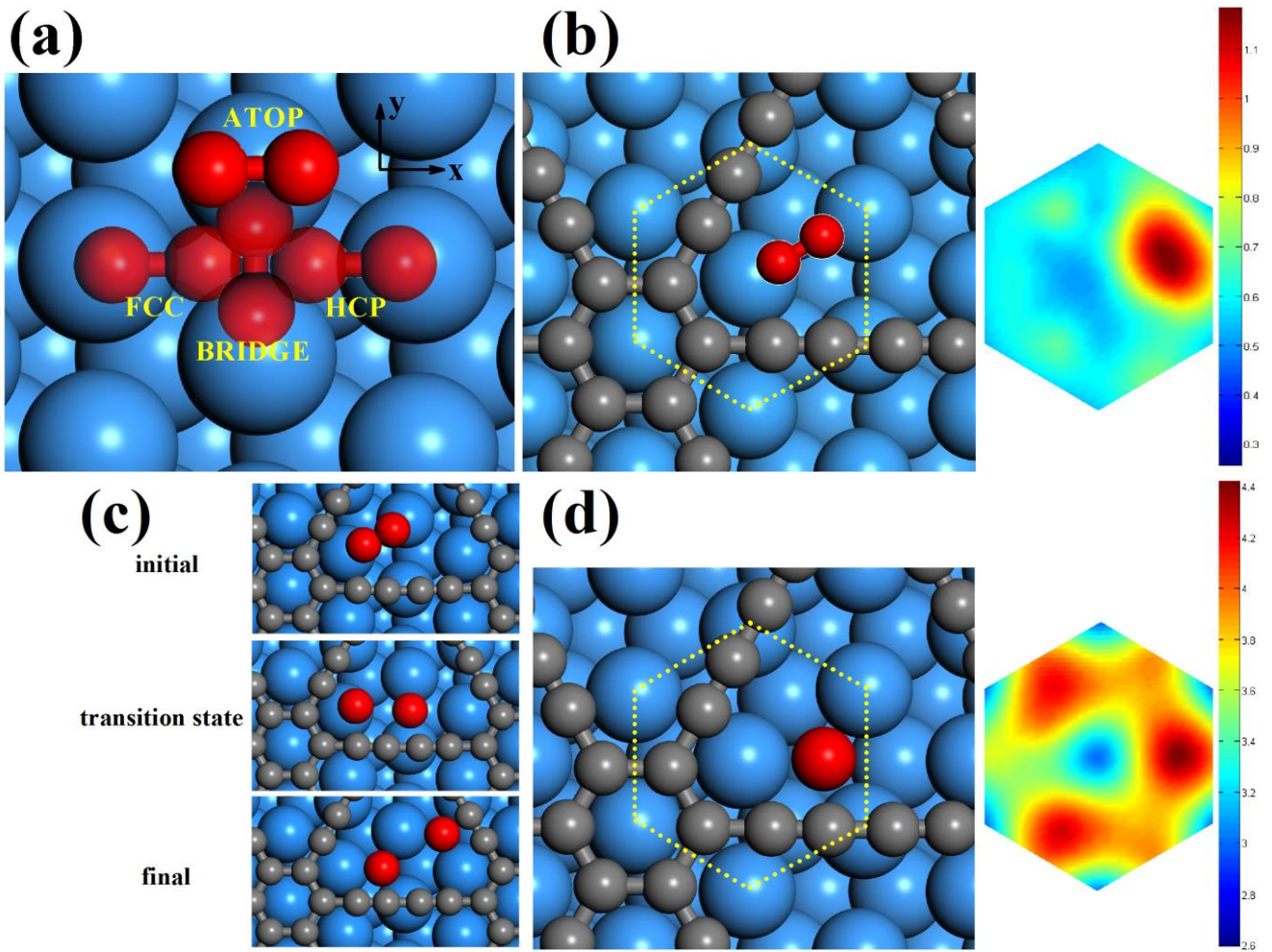

Fig. 2 (**a**) Schematic of FCC, HCP, BRIDGE and ATOP adsorption sites of $O_2$ on Pt(111) surface. The left panel of (**b**)/(**d**) shows the most stable site for $O_2$/O underneath graphdiyne, and the color map represents the PES of $O_2$/O in the irreducible area (enclosed by dashed lines), respectively. The colorbars show the binding energies in eV. (**c**) $O_2$ dissociation on graphdiyne-covered Pt(111) surface. Pt, C and O atoms are represented in blue, gray and red, respectively.

To investigate $O_2$ adsorption on graphdiyne-covered Pt(111) surface, the potential energy map of $O_2$ molecule was plotted in the irreducible area shown by the hexagonal dotted lines in Fig.2 (b). Geometry relaxations and calculations of binding energy $E_b$ (including ZPE corrections) were performed with the centroid of $O_2$ fixed on every FCC/HCP/BRIDGE/ATOP site underneath the graphdiyne cover. All other positions outside can be mapped into the irreducible area according to the symmetry. In the irreducible area, the potential energy surface (PES) was plotted using data interpolation of binding energies at all FCC, HCP, BRIDGE and ATOP sites. The color map of $O_2$ PES underneath the graphdiyne cover is shown in the right of Fig. 2(b). The BRIDGE site in the triangular C acetylenic ring was found to be the most stable adsorption site (see the left panel of Fig. 2(b)) with a binding energy of $E_b$ = 1.142 eV, and the neighboring FCC site is the second most stable with a binding energy of $E_b$ = 1.138 eV (the red area in the right panel of Fig. 2(b)). With largest binding energies, the BRIDGE and FCC sites in the acetylenic triangle are thermodynamically favorable for $O_2$ adsorption. When $O_2$ gets close to the C skeleton of graphdiyne, the repulsion between graphdiyne and $O_2$ causes the decrease of binding energy. In the right panel of Fig. 2(b), we can see a large area with lower binding energy (about $E_b$ = 0.6~0.7 eV, the blue area in the right panel of Fig. 2(b)) due to the steric hindrance of graphdiyne sheet. This area is energetically unfavorable for $O_2$ molecule to access. According to the above results, the $O_2$ adsorption mainly happens in the separated area surrounded by graphdiyne acetylenic rings. The binding energy of $O_2$ on graphdiyne-covered Pt(111) surface is larger than on pristine Pt(111) surface, which is advantageous to the pre-adsorption of $O_2$.

The barrier and standard free energy of activation of $O_2$ dissociation on graphdiyne-covered Pt(111) surface were found to be $E_a$=0.34 eV and $\Delta G^{\neq}$ = 0.28 eV, respectively, which are lower than $E_a$=0.38 eV and $\Delta G^{\neq}$ = 0.33 eV of the dissociation on pristine Pt(111) surface. At $T$ = 300 K and pressure $p$ = 1 atm, the rate constant of $O_2$ dissociation on graphdiyne-covered Pt(111) surface is $r = \frac{kT}{h}\exp(-\Delta G^{\neq}/kT) = 1.2 \times 10^8$ s$^{-1}$, which is about seven times of $r$ = 1.8×10$^7$ s$^{-1}$ on pristine Pt(111) surface. In the primary step of $O_2$ dissociation (Fig. 2(c)), one of the O atoms goes across the neighboring HCP site and reaches the nearby FCC site with elongating O-O bond length. The standard free energy change from $O_2$ (the initial configuration in Fig. 2(c)) to two O atoms located at nearest FCC sites (the final configuration in Fig. 2(c)), i.e. the coadsorption state of O atoms, is $\Delta G$ = -1.70 eV. Then, two O atoms may migrate and become totally separated. The standard free energy change $\Delta G$ = -1.91 eV from $O_2$ to two totally separated O atoms is larger than on pristine Pt(111) surface ($\Delta G$ = -1.62 eV). Overall, the above results sufficiently demonstrate that the $O_2$ dissociation on graphdiyne-covered Pt(111) surface is more thermodynamically favorable than on pristine Pt(111) surface, which provides a prerequisite for CO oxidation.

To plot the PES of O atom on graphdiyne-covered Pt(111) surface, geometry relaxations and binding energy calculations (including ZPE corrections) were performed with the O atom put on every FCC/HCP/BRIDGE/ATOP site. The interpolated PES is shown in the right panel of Fig. 2(d). According to the results, the binding energy of O atom on graphdiyne-covered Pt(111) surface (3.5~4.3 eV) is close to that on pristine Pt(111) surface (4.0~4.5 eV). The FCC site in the triangular C acetylenic ring was found to be the most stable adsorption site (see the left panel of Fig. 2(d)) with a binding energy of $E_b$ = 4.34 eV. The binding energies on FCC sites (4.11~4.34 eV) are generally higher than on HCP sites (3.51~3.76 eV). Nevertheless, the binding energies on the FCC sites underneath the C skeleton of graphdiyne (4.11~4.16 eV) are lower than on the most stable FCC site (4.34 eV) due to the steric hindrance. According to the above results, the energy difference between FCC and HCP sites is less than 0.9 eV. At room temperature, the migration of O atom on Pt surface might be possible.

3.3 *CO adsorption on graphdiyne-covered* Pt(111) *surface*

To investigate CO adsorption on graphdiyne-covered Pt(111) surface, geometry relaxations and calculations of binding energy (including ZPE corrections) were performed with CO fixed on every FCC/HCP/BRIDGE/ATOP site, and the PES of CO was plotted in the irreducible area shown by the hexagonal dotted lines in Fig. 3 (a). CO prefers to be perpendicular to the Pt surface, with the C atom down. On graphdiyne-covered Pt(111), the O atom of standing CO molecule gets close to the graphdiyne sheet, causing more steric hindrance than the lying $O_2$ molecule. In the region of triangular C acetylenic ring, CO suffers less hindrance and obtains larger binding energy. The ATOP site in the middle of triangular C acetylenic ring (see the left panel and the box in Fig. 3(a)) was found to be the most stable adsorption site with a binding energy $E_b$ = 1.56 eV, which is slightly lower than $E_b$ = 1.89 eV on pristine Pt(111) surface. The binding energies on FCC sites are much lower, which are in the range of 0.38~0.45 eV. On other adsorption sites, the binding energies are even below 0.35 eV. So, the center ATOP site is much more favorable than other sites in thermodynamics. At $T$ = 300 K and pressure $p$ = 1 atm, the standard free energy change of CO adsorption is $\Delta G$ = -1.04 eV. The energy difference between the most stable ATOP site and the neighboring FCC sites is 1.11 eV, and the energy differences between all the neighboring FCC and HCP sites are in the range of 0.08~0.15 eV. Since the energy difference between these adsorption sites are around 1 eV, the CO thermal migration on graphdiyne-covered Pt(111) surface may be possible.

| CO migration step | 1 | 2 | 3 | 4 | 5 |
|---|---|---|---|---|---|
| potential barrier $E_a$ (eV) | 1.17 | 0.26 | 0.10 | 0.06 | 0.05 |

Table 1 The potential barriers of CO migration steps **1~5** in Fig. 3(**b**).

| $O_2$ migration step | 1 | 2 | 3 | 4 | 5 | 6 |
|---|---|---|---|---|---|---|
| potential barrier $E_a$ (eV) | 0.66 | 0.39 | 0.73 | 0.08 | 0.81 | 0.23 |

Table 2 The potential barriers of $O_2$ migration steps **1~6** in Fig. 3(**c**).

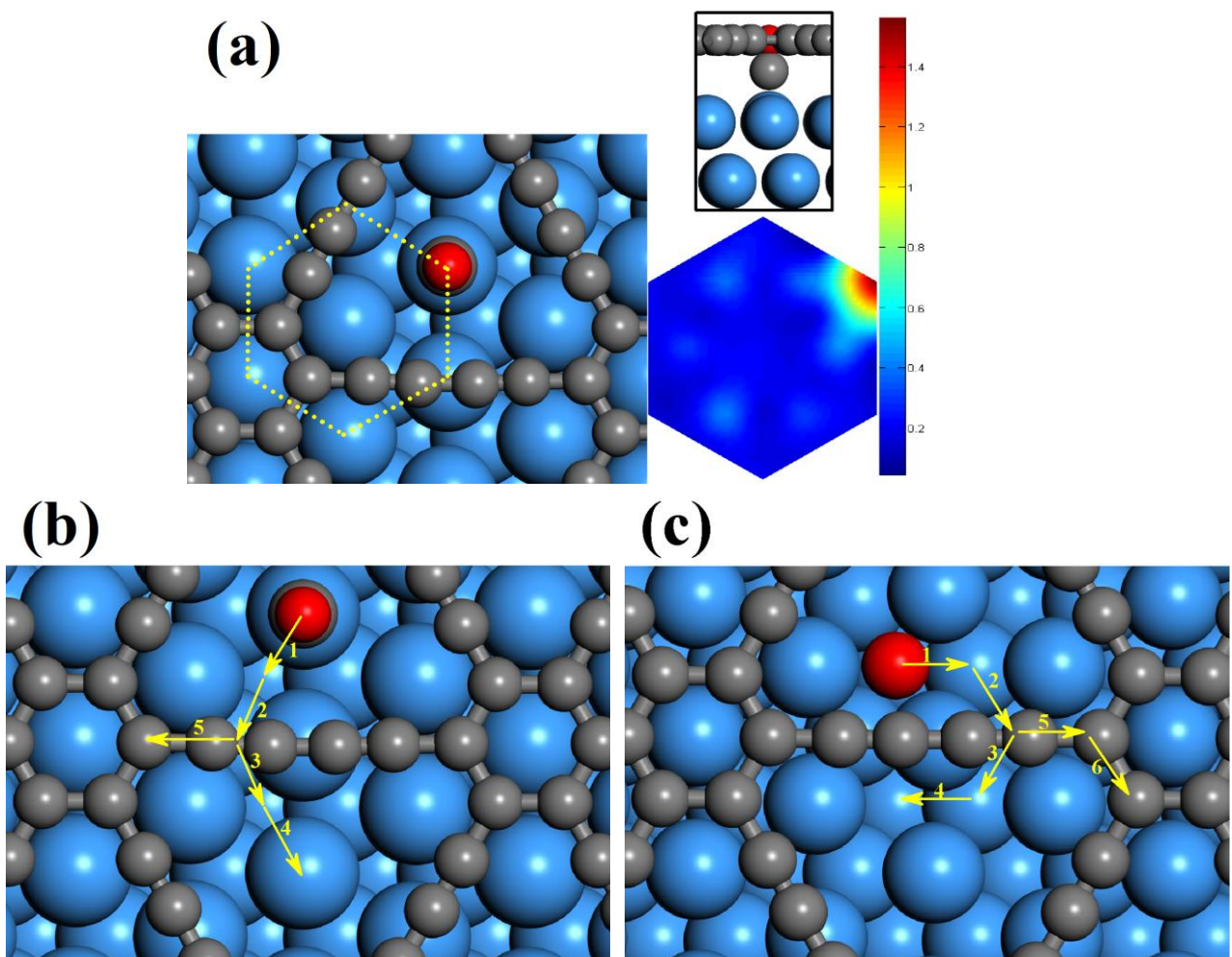

Fig. 3 (**a**) The top/side views of the most stable site for CO on graphdiyne-covered Pt(111) surface are shown in the left panel/the box, respectively. The color map represents the PES of CO in the irreducible area (enclosed by dashed lines). (**b**) Migration paths of CO on graphdiyne-covered Pt(111) surface. (**c**) Migration paths of O on graphdiyne-covered Pt(111) surface. Pt, C and O atoms are represented in blue, gray and red, respectively.

3.4 *CO and* O *migration on graphdiyne-covered* Pt(111) *surface*

To further demonstrate the existence of prerequisite for CO oxidation, we investigated the migration of CO and O on graphdiyne-covered Pt(111) surface. Potential barrier calculations indicate that CO and O could migrate and encounter each other at room temperature. Since the CO migration process on graphdiyne-covered Pt(111) surface is complex with many different adsorption sites and migration steps, we only investigated some typical steps in the calculations. The arrows of **1~4** in Fig. 3(b) presents the shortest CO migration path from one ATOP site to another, with corresponding potential barriers shown in Table 1. The step **1** from the most stable ATOP site to the neighboring FCC site has the highest barrier $E_a$ = 1.17 eV. These following steps **2~4** are of much lower barriers than step **1**, indicating that step **1** is the rate-determining step. At 300 K, the standard free energy of activation of step **1** is $\Delta G^{\neq}$ = 1.11 eV, and the corresponding rate constant is $r = \frac{kT}{h}\exp(-\Delta G^{\neq}/kT) = 1.4 \times 10^{-6}$ s$^{-1}$. Such rate is too low for CO to migrate at room temperature. At $T$ = 400 K, the rate constant of step **1** is $r = 6.5 \times 10^{-2}$ s$^{-1}$, which is passable for the reaction to proceed.

Via step **1**, CO arrives at the neighboring FCC site. Then, the CO molecule may go across the HCP site under the acetylenic chain via step **2~4**, and arrive at another ATOP site, with corresponding potential barriers $E_a$ below 0.3 eV (see Table 1). CO may also return to the ATOP site via the inverse process of step **1**, or go to other HCP sites. The potential barrier of step **2** (the inverse process of step **1**) is $E_a$ = 0.26 (0.06) eV, respectively. Since the barriers are very low, these processes could easily happen at room temperature. Besides step **2~4**, other paths (e.g. step **5**) may be also thermodynamically possible. Because of the complexity of calculating all the migration paths, here we only investigate the shortest route from one ATOP site to another, showing the possibility of CO migration.

Summing up the above analysis, CO migration on graphdiyne-covered Pt(111) surface is difficult at room temperature due to too slow the rate-determining step **1**. At $T$ = 400 K or above, the rate of step **1** is passable and CO migration becomes possible. Once overcoming the barrier of step **1**, the barriers following steps are much lower and CO would move from one ATOP site to another.

Then, the migration of O on graphdiyne-covered Pt(111) surface was investigated, finding lower barriers than CO migration. We inferred that O migration is possible at room temperature. The arrows of **1~4** in Fig. 3(c) presents the shortest O migration path from one most stable FCC site to another, with corresponding potential barriers shown in Table 2. With the highest barrier $E_a$ = 0.66 eV, step **1** was considered as the rate-determining step. At 300 K, the standard free energy of activation of step **1** is $\Delta G^{\neq}$ = 0.60 eV, and the corresponding rate constant is $r = \frac{kT}{h}\exp(-\Delta G^{\neq}/kT) =$ $5.2 \times 10^2$ s$^{-1}$, which is large enough for step **1** to proceed at room temperature. The following steps **2~4** are with lower barriers (below 0.4 eV, see Table 2). Another migration path, i.e. steps **5** and **6**, are less possible because the barrier $E_a = 0.81$ eV of step **5** is high than step **1**. Overall, with lower barriers than CO, O migration may be possible with high rate at room temperature, providing opportunities for O to encounter CO and perform the oxidation reaction.

3.5 *CO oxidation on graphdiyne-covered* Pt(111) *surface*

With adsorbed CO molecules and dissociated O atoms on graphdiyne-covered Pt(111) surface, the reaction O + CO → $CO_2$ happens when O atom migrates and meets CO molecule. As discussed in Sec. 3.4, with much lower barrier than CO, O atom is easier to migrate at room temperature. Here, we considered the process of one O atom attacking CO. First, for a CO molecule located at the most stable ATOP site, one O atom approaches it and arrives at the nearest HCP (or FCC) site underneath the C acetylenic chain, forming O-HCP (or O-FCC) coadsorption state (the left panel of Fig. 4(**a**)). At $T = 300$ K, the standard free energy change of forming the O-HCP (O-FCC) state is $\Delta G = 0.82$ (0.22) eV, respectively. Second, the O atom climbs over the barrier, binding with CO and forming a $CO_2$ molecule. Starting from the O-HCP (O-FCC) state, the standard free energy of activation of O + CO reaction is $\Delta G^{\neq} = 0.22$ (2.24) eV. The initial, transition and final states of O-HCP route are shown in Fig. 4(**b**). The very high barrier of O-FCC route indicates that this route is unfavorable. The right panel of Fig. 4(**a**) portrays the free energy of two different reaction paths. In the thermodynamically favorable O-HCP route, the process of forming O-HCP coadsorption state is the rate-determining step, and the total free energy barrier $\Delta G^{\neq} = 0.82 + 0.22 = 1.04$ eV is lower than the calculated total free energy barrier $\Delta G^{\neq} = 1.09$ eV for pristine Pt(111). Finally, the formed $CO_2$ desorbs from graphdiyne-covered Pt(111) surface, with a desorption free energy change $\Delta G = -0.117$ eV at $T = 300$ K and pressure $p = 1$ atm. The very weak binding energy of $CO_2$ ($E_b = 0.10$ eV) and the entropy increase of $CO_2$ into gas phase leads to negative $\Delta G$, which is advantageous for CO$_2$ desorption. Overall, the above results indicate the possibility of O + CO → CO$_2$ on graphdiyne-covered Pt(111) surface. The standard free energy change of the reaction is calculated to be Δ$G$ = -1.01 eV, which is much more negative than Δ$G$ = -0.52 eV on pristine Pt(111) surface. So, the reaction O + CO → CO$_2$ on graphdiyne-covered Pt(111) surface is thermodynamically favorable and could be performed at room temperature.

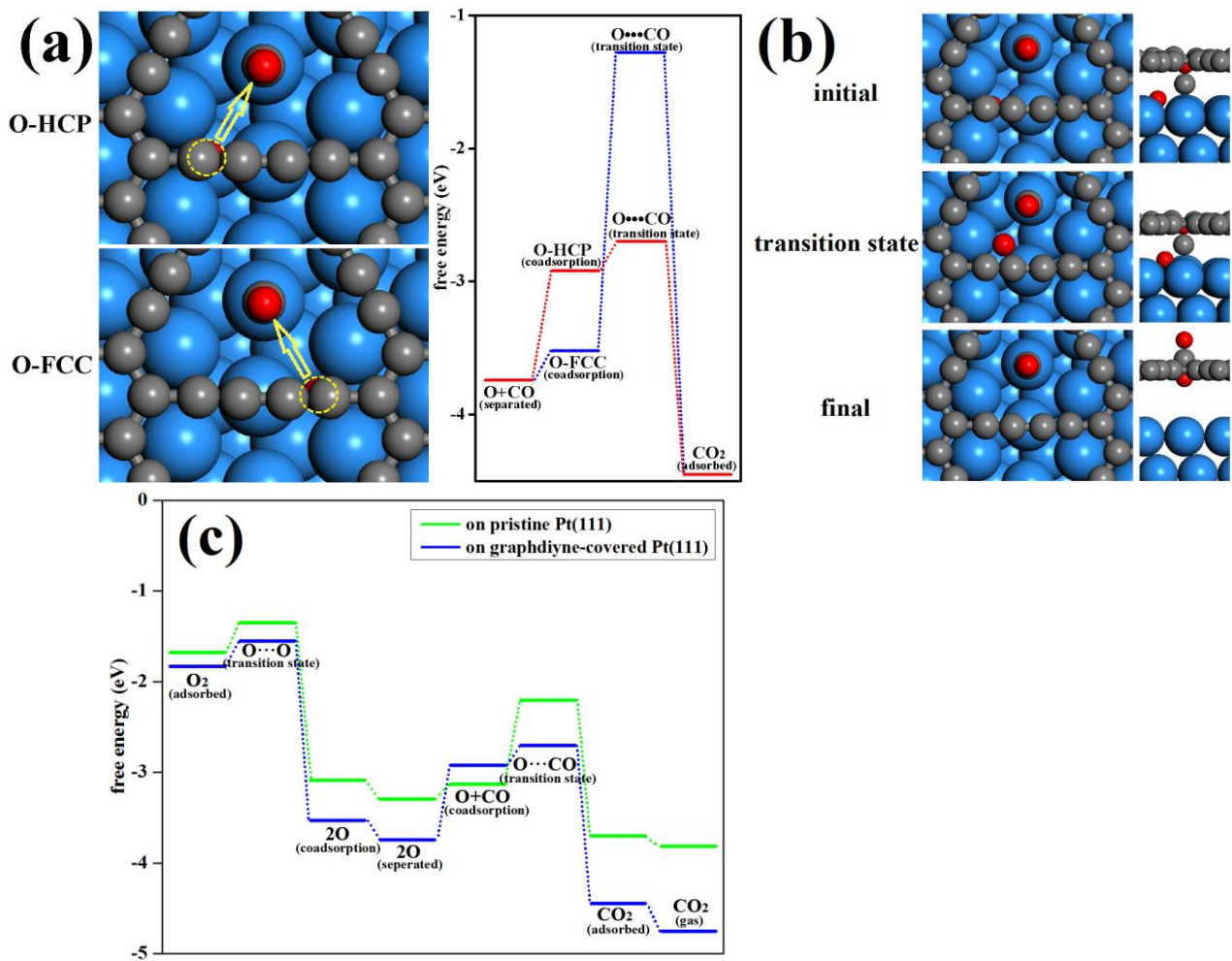

Fig. 4 (**a**) The left panel presents the O-HCP and O-FCC coadsorption states of O and CO. The dashed circles marks the O atoms underneath the C acetylenic chain. The arrows shows the attack directions of O atoms. The right panel shows the free energy chart of two different reaction paths for O + CO → CO$_2$ on graphdiyne-covered Pt(111) surface. (**b**) The O-HCP path of CO + O → CO$_2$ on graphdiyne-covered Pt(111) surface, with the top and side views of initial, transition and final states shown. Pt, C and O atoms are represented in blue, gray and red, respectively. (**c**) The free energy chart of total reaction paths on pristine (green) and graphdiyne-covered (blue) Pt(111).

Summarizing the above discussion about O$_2$ dissociation (Sec. 3.2), O migration (Sec. 3.4) and O + CO reaction, the total reaction path on graphdiyne-covered Pt(111) surface is more energetically favorable than on pristine Pt(111). The free energy landscape from $O_2$ dissociation, CO oxidation to $CO_2$ production is plotted in Fig. 4(**c**). It can be seen that the reaction path on graphdiyne-covered Pt(111) surface generally has lower free energy than on pristine Pt(111). On graphdiyne-covered Pt(111) surface, the lower $\Delta G$ of $O_2$ adsorption and dissociation promotes the reaction towards the positive direction better than on pristine Pt(111). Next, O + CO step (including O + CO coadsorption and reaction substeps) plays a role of determining the overall rate. On graphdiyne-covered Pt(111) surface, the O + CO coadsorption substep with $\Delta G = 0.82$ eV is the largest free energy barrier. On pristine Pt(111), the O + CO reaction substep with $\Delta G^{\neq} = 0.92$ eV plays a role of the largest free energy barrier, which is higher than on graphdiyne-covered Pt(111) surface. Furthermore, the $\Delta G$ of O + CO → $CO_2$ reaction is more negative on graphdiyne-covered Pt(111) surface. Overall, graphdiyne-covered Pt(111) surface has the advantage on free energy to catalyze CO oxidation.

## 4. Conclusions

In this work, graphdiyne-covered Pt(111) surface was taken as a model to study the superiority of graphdiyne in two-dimensional confined catalysis and unveil the catalysis mechanism. Graphdiyne overlayer on Pt(111) surface was proved to be excellent catalyst to promote CO oxidation at room temperature. The comparison of calculation results exhibits the better catalytic ability of graphdiyne-covered Pt(111) than pristine Pt(111). The CO oxidation includes CO and $O_2$ adsorption, $O_2$ dissociation, O atom migration and O + CO → $CO_2$ step. On graphdiyne-covered Pt(111) surface, the adsorbed CO and $O_2$ molecules are limited in the C acetylenic ring due to the steric hindrance. With less hindrance and lower migration barriers, the dissociated O atoms are free to move on graphdiyne-covered Pt(111) surface at room temperature. The wandering O atom meets CO and forms $CO_2$. Finally, with very low binding energy, $CO_2$ is easy to leave the Pt surface.

Using free energy calculations, the free energy landscape of the whole reaction was presented to demonstrate the superiority of graphdiyne-covered Pt(111) to pristine Pt(111). At room temperature, the free energy changes of $O_2$ dissociation and the $O + CO \rightarrow CO_2$ step on graphdiyne-covered Pt(111) are more negative than on pristine Pt(111). Furthermore, the barriers of $O_2$ dissociation and the $O + CO \rightarrow CO_2$ step on graphdiyne-covered Pt(111) are lower than on pristine Pt(111). The advantages in free energy changes and potential barriers of the key reaction steps make graphdiyne-covered Pt(111) superior to pristine Pt(111) in catalysis. In addition, the acetylenic triangle pores of graphdiyne play a role of natural CO and $O_2$ entrance and $CO_2$ exit. By contrast with graphene overlayer using fractures as molecular tunnels, graphdiyne overlayer could be more excellent in two-dimensional confined catalysis [33], and would improve the heterogeneous catalysis ability of transition metal surface.

**Acknowledgements**

This work was supported by the National Natural Science Foundation of China under Grant No. 11304239, and the Fundamental Research Funds for the Central Universities.